\documentclass[showpacs,preprintnumbers,amsmath,amssymb,twocolumn,superscriptaddress,prb]{revtex4}
\usepackage{graphicx}
\usepackage{amsfonts}
\usepackage{amsmath, amsthm, amssymb}
\usepackage{dsfont}
\usepackage{times}
\usepackage{subfigure}

\newcommand{\e}[1]{\mathrm{e}^{#1}}

\newcommand{\bq}{\begin{equation}}
\newcommand{\eq}{\end{equation}}

\newcommand{\g}{\underline{\gamma}}
\newcommand{\gt}{\underline{\tilde{\gamma}}}
\newcommand{\N}{\underline{\mathcal{N}}}
\newcommand{\Nt}{\underline{\tilde{\mathcal{N}}}}

\newcommand{\eg}{\textit{e.g. }}
\newcommand{\etal}{\emph{et al.}}
\def\i{\mathrm{i}}

\begin{document}
\title[Phase-controlled proximity-effect in ferromagnetic Josephson junctions: calculation of DOS and electronic specific heat]{Phase-controlled proximity-effect in ferromagnetic Josephson junctions: calculation of DOS and electronic specific heat}

\author{Mohammad Alidoust}
\affiliation{Department of Physics, Faculty of Sciences,
University of Isfahan, Hezar Jerib Avenue, Isfahan 81746-73441,
Iran}

\author{Jacob Linder}
\affiliation{Department of Physics, Norwegian University of
Science and Technology, N-7491 Trondheim, Norway}

\author{Gholamreza Rashedi}
\affiliation{Department of Physics, Faculty of Sciences,
University of Isfahan, Hezar Jerib Avenue, Isfahan 81746-73441,
Iran}

\author{Asle Sudb{\o}}
\affiliation{Department of Physics, Norwegian University of
Science and Technology, N-7491 Trondheim, Norway}

\date{Received \today}

\begin{abstract}
We study the thermodynamic properties of a dirty ferromagnetic
S$\mid$F$\mid$S Josephson junction with $s$-wave superconducting
leads in the low-temperature regime. We employ a full numerical
solution with a set of realistic parameters and boundary
conditions, considering both a uniform and non-uniform exchange
field in the form of a Bloch domain wall ferromagnetic layer. The
influence of spin-active interfaces is incorporated via a
microscopic approach. We mainly focus on how the electronic
specific heat and density of states (DOS) of such a system is
affected by the \textit{proximity effect}, which may be tuned via
the superconducting phase difference. Our main result is that it
is possible to \textit{strongly modify the electronic specific
heat} of the system by changing the phase difference between the
two superconducting leads from $0$ up to nearly $\pi$ at low
temperatures. An enhancement of the specific heat will occur for
small values $h\simeq\Delta$ of the exchange field, while for
large values of $h$ the specific heat is suppressed by increasing
the phase difference between the superconducting leads. These
results are all explained in terms of the proximity-altered DOS in
the ferromagnetic region, and we discuss possible methods for
experimental detection of the predicted effect.
\end{abstract}

 \maketitle

\section{Introduction}
In recent years, due to the important role of hybrid structures with superconducting and magnetic layers in vital circuit-elements like transistors and high-resolution devices like detectors, such structures has attracted
much attention from the research community. In this way, several interesting
phenomena of such systems have been predicted both in the dirty and
clean limits of transport, and subsequently been observed in experiments:
non-monotonic dependence of the critical temperature $T_{c}$ in
S$\mid$F hybrid structures on the thickness of F
layer,\cite{muehge,radovic,jiang,buzdin1,tagirov} the $0$-$\pi$
transitions in S$\mid$F$\mid$S
junctions,\cite{ryazanov,buzdin2,buzdin3}, and the appearance of odd-frequency pairing correlations \cite{bergeret_prl_01, keizer_nature_06} just to mention a few. The main cause of the appearance of
these interesting phenomena is the \textit{proximity effect} between the
superconductor and the ferromagnet, where Cooper pairs leak from the superconducting side to the ferromagnetic layer. S$\mid$F$\mid$S junctions are routinely fabricated by experimentalists these days, and currently such hybrid structures are
intensely investigated due to their potential both in terms of
functionality\cite{lee_book_01} and novel fundamental physics that
may be explored.\cite{bergeret_rmp_05,buzdin4}
\par
In the context of applications, studies of the
thermodynamic properties of superconductors have mostly focused on
electron cooling properties \cite{giazotto_rmp_06}, although
recently the influence of the proximity effect on the entropy
production in a non-magnetic Josephson junction was investigated
\cite{rabani}. A key to understanding the thermodynamic properties
of a system is the behavior of the DOS near Fermi level, and any
control parameter that can adjust the DOS in an efficient and
well-defined manner would offer significant advantages with
respect to tailoring desired thermodynamic properties.
\par
Very recently, it has been studied numerically
\cite{hammer_prb_07} and demonstrated experimentally
\cite{sueur_prl_08} how the density of states (DOS) may be altered
controllably in such structures by creating a non-magnetic
S$\mid$N$\mid$S Josephson junction and generating a supercurrent.
For a ferromagnetic Josephson junction, however, it remains to be
clarified precisely how the DOS is influenced by the phase
difference in the full proximity effect regime. A new feature
which is expected to come into play for ferromagnetic Josephson
junctions is the presence of odd-frequency superconducting
correlations, which can induce a qualitative shift in the DOS from
a low-energy minigap-structure \cite{mcmillan_pr_68} to an
enhancement \cite{oddfreq}.

\par
In this paper, we show how it is possible to obtain a \textit{huge
enhancement of the specific heat} of a
ferromagnetic Josephson junction at low temperatures, simply by tuning the
superconducting phase difference by means of either a current or
an external magnetic flux in a SQUID-like geometry. We demonstrate
explicitly how the predicted effect occurs for a set of realistic
experimental parameters, and how it persists even in the presence
of an inhomogeneous magnetization texture such as a Bloch domain
wall in the ferromagnet. We find that the enhancement of the specific
heat is strongest for exchange fields $h$ comparable in magnitude to the superconducting gap $\Delta$, i.e. $h\simeq\Delta$, whereas for higher exchange fields the effect eventually vanishes.
Our findings can be verified
experimentally by using calorimetry techniques or high-resolution
thermometry, \cite{zassenhaus_jltp_98} and could have potential
applications in devices utilizing an active tuning of the
thermodynamic properties of nanoscale conductors. We underline
that while it is well-known that the DOS in a Josephson junction
is sensitive to the phase difference, our main result pertains to
\textit{the manner in which the DOS varies and the resulting
consequences for the electronic specific heat of the junction}.

\begin{figure}
\includegraphics[width=7cm]{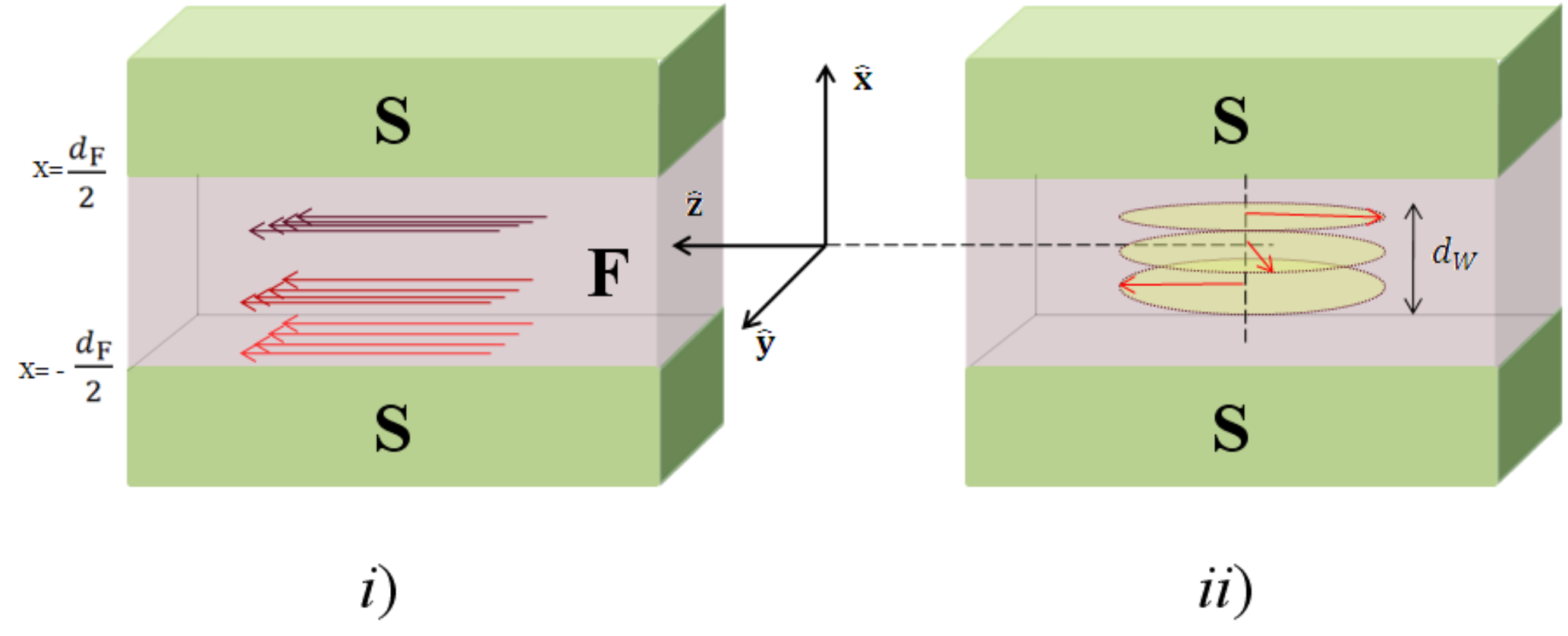}
\caption{\label{fig:model} (Color online) The
S$\mid$F$\mid$S junction with \textit{i)} uniform ferromagnet in the $\hat{z}$
direction and \textit{ii)} with a Bloch domain wall in the ferromagnet.
The magnetization texture for the N\'{e}el wall would be obtained by
replacing the $x$-component of the magnetization with an
$y$-component in case \textit{ii)}. The red arrows show the
magnetic moments in the F layer.}
\end{figure}

\section{Theory}

To investigate the physical properties of the S$\mid$F$\mid$S
Josephson junction, one alternative is to solve the quasiclassical Eilenberger equation \cite{eilenberger,belzig_review} to obtain the Green's functions. In
the diffusive limit, the Eilenberger equation reduces to a simpler set of equations known as the Usadel equations \cite{usadel}. For numerical purposes, it is convenient to use a parameterization method for the Green's
functions. One approach for the parameterization is the
$\theta$-parametrization\cite{ivanov} as follows:
\begin{align}
\hat{g} &= \begin{pmatrix}
M_0c\underline{\sigma_0} + (\boldsymbol{M}\cdot\underline{\boldsymbol{\sigma}})s & \underline{\rho}^+ \notag\\
\underline{\rho}^- & -M_0c\underline{\sigma_0} -
(\boldsymbol{M}\cdot\underline{\boldsymbol{\sigma}})^*s
\end{pmatrix},\notag\\
\underline{\rho}^\pm &=
c[\i(M_z\underline{\sigma_2}-M_y\underline{\sigma_3})\pm
M_x\underline{\sigma_0}] \pm M_0\underline{\sigma_1}s,
\end{align}
where $\underline{\sigma_j}$ are the identity $(j=0)$ and Pauli
$(j=1,2,3)$ matrices, and \begin{align}
\underline{\boldsymbol{\sigma}} = (\underline{\sigma_1},
\underline{\sigma_2}, \underline{\sigma_3}).
\end{align}
Also, $s\equiv \sinh(\theta)$ and $c\equiv \cosh(\theta)$. The
Green's function is then completely determined by the complex
functions $\theta$, $M_0$, and $\boldsymbol{M}$ with the
additional constraint $M_0^2 -\boldsymbol{M}^2=1$ in order to
satisfy $\hat{g}^2=\hat{1}$. The other approach for a parameterization of
the Green's functions is the
Ricatti-parameterization\cite{Schopohl,Konstandin}. We found that
for our purposes in this paper, \textit{i.e.} a full numerical
investigation of the density of states and consequently the thermodynamic
properties of a diffusive S$\mid$F$\mid$S junctions, this parametrization is much more numerically stable than the $\theta$-parameterization. The
Ricatti-parameterization can read as
follow:\cite{Konstandin,hammer_prb_07}
\begin{align}\label{eq:g}
\hat{g} &= \begin{pmatrix}
\N(\underline{1}-\g\gt) & 2\N\g \\
2\Nt\gt & \Nt(-\underline{1} + \gt\g) \\
\end{pmatrix}.
\end{align}
Here, $\hat{g}^2=\hat{1}$ since
\begin{align}
\N=(1+\g\gt)^{-1}\; \Nt = (1+\gt\g)^{-1}.
\end{align}
We use $\underline{\ldots}$ for $2\times2$ matrices and
$\hat{\ldots}$ for $4\times4$ matrices. In order to calculate the
Green's function $\hat{g}$, we need to solve the Usadel equation
\cite{usadel} with appropriate boundary conditions at $x=-d_F/2$
and $x=d_F/2$. We introduce the superconducting coherence length
as $\xi_S = \sqrt{D_S/\Delta_0}$. Following the notation of Ref.
\cite{linder_prb_09}, the Usadel equation reads
\begin{align}\label{eq:usadel}
D\partial(\hat{g}\partial\hat{g}) + \i[E\hat{\rho}_3 +
\text{diag}
[\boldsymbol{h}\cdot\underline{\boldsymbol{\sigma}},(\boldsymbol{h}\cdot
\underline{\boldsymbol{\sigma}})^\mathcal{T}], \hat{g}]=0,
\end{align}
and we employ the following realistic boundary conditions for all
our computations in this paper: \cite{Hernando}
\begin{align}\label{eq:bc}
2\zeta d_F\hat{g} \partial \hat{g} = [\hat{g}_\text{BCS}(\phi),
\hat{g}] + \i (G_S/G_T) [\text{diag}(\underline{\tau_3},
\underline{\tau_3}), \hat{g}]
\end{align}
at $x=-d_F/2$. Here, $\partial \equiv \frac{\partial}{\partial x}$
and we defined $\zeta=R_B/R_F$ as the ratio between the resistance
of the barrier region and the resistance in the ferromagnetic
film. The barrier conductance is given by $G_T$, whereas the
parameter $G_S$ describes the spin-dependent interfacial phase-shifts (spin-DIPS) taking place at the F side
of the interface where the magnetization is assumed to lie in the $yz$-plane, being parallel to
the $z$-axis at the interfaces. The boundary condition at $x=d_F/2$ is obtained by
letting $G_S \to (-\tilde{G}_S)$ and $\hat{g}_\text{BCS}(\phi) \to
[-\hat{g}_\text{BCS} (-\phi)]$ in Eq. (\ref{eq:bc}), where
\begin{align}
\g_\text{BCS}(\phi) &= \i\underline{\tau_2}s/
(1+c)\e{\i\phi/2},\notag\\
\gt_\text{BCS}(\phi) &= \g_\text{BCS}(\phi)\e{-\i\phi}.
\end{align}
Above, $\tilde{G}_S$ is allowed to be different from $G_S$ in
general. For instance, if the exchange field has opposite
direction at the two interfaces due to the presence of a domain
wall, one finds $\tilde{G}_S=-G_S$. The total superconducting
phase difference is $\phi$, and we have defined
$s=\sinh(\vartheta), c=\cosh(\vartheta)$ with
$\vartheta=\text{atanh}(\Delta_0/E)$ using $\Delta_0$ as
the superconducting gap. Note that we use the bulk solution in the
superconducting region, which is a good approximation when
assuming that the superconducting region is much less disordered
than the ferromagnet and when the interface transparency is small,
as considered here. Effectively, the inverse proximity effect is thus ignored. We use units such that $\hbar=k_B=1$.

The values of $G_S$ and $G_T$ may be calculated explicitly from a
microscopic model, which allows one to characterize the
transmission $\{t_{n,\sigma}^j\}$ and reflection amplitudes
$\{r_{n,\sigma}^j\}$ on the $j\in\{S,F\}$ side. Under the
assumption of tunnel contacts and a weak ferromagnet, one obtains
with a Dirac-like barrier model\cite{Cottet_1,Cottet_2, Hernando}
\begin{align}
G_T = G_Q\sum_n T_n,\; G_S = 2G_Q\sum_n\Big( \rho_n^F -
\frac{4\tau_n^S}{T_n}\Big)
\end{align}
upon defining $T_n = \sum_\sigma |t_{n,\sigma}^S|^2$, $ \rho_n^F =
\text{Im}\{r_{n,\uparrow}^F (r_{n,\downarrow}^F)^*\}$ and
$\tau_n^S = \text{Im}\{t_{n,\uparrow}^S (t_{n,\downarrow}^S)^*\}$
and also for simplicity, we assume that the interface is
characterized by $N$ identical scattering channels and
consequently omit the subscript '$n$' in the summations.

\begin{figure*}
\includegraphics[width=18cm]{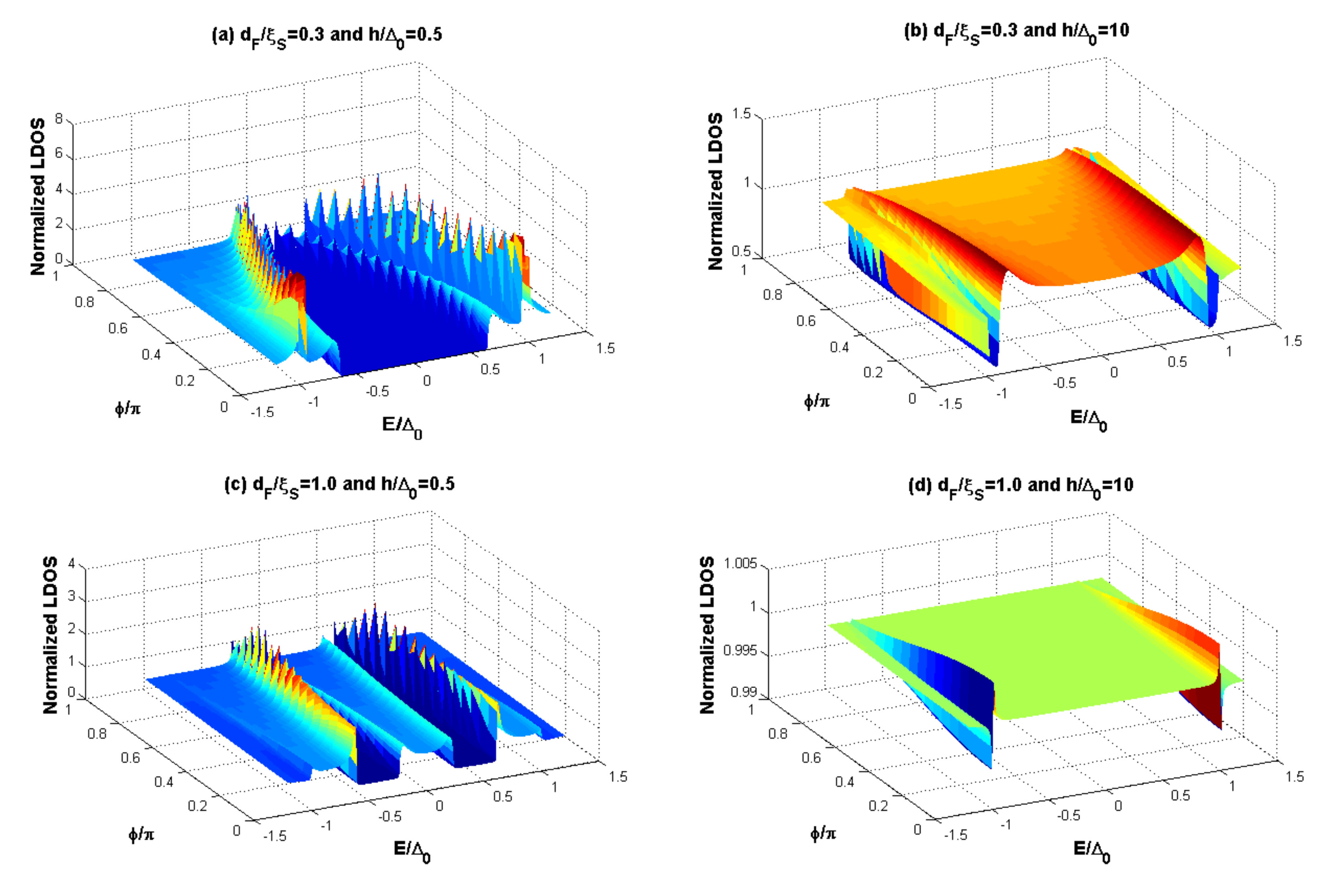}
\caption{\label{fig:DOS1} (Color online) The
normalized local density of states vs. energy and phase difference
between two $s$-wave superconducting leads for a uniform
magnetization texture. We have presented several values of the thickness of
ferromagnetic layer $d_{F}/\xi_{S}$ and exchange field
$h/\Delta_{0}$, and the DOS is evaluated in the middle of the junction, \textit{i.e.} $x=0$.}
\end{figure*}

To find the specific heat of the system, we need to calculate the
local density of states normalized against its normal-state value
\begin{align}\label{eq:DOS}
N(x,E,T,\phi)=\text{Tr}(\text{Re}[\N(\underline{1}-\g\gt)])/2.
\end{align}
We assume that the S electrodes are not influenced by the
proximity effect, the total electronic specific heat
($C_{\text{tot}}$) of the SFS junction can be determined by
\begin{align}\label{eq:HC_tot}
C_{\text{tot}} (T,\phi)=C_\text{F}(T,\phi)+C_\text{S}(T).
\end{align}
Here, $C_\text{S}(T)$ is the specific heat
 of superconducting plates while
\begin{align}\label{eq:HC}
C_\text{F}(T,\phi)=T\partial{S_\text{F}}(T,\phi)/\partial{T},
\end{align}
is the specific heat of the ferromagnetic part of the junction.
The entropy of the ferromagnet layer in the proximity system can
be
obtained from
\begin{align}\label{eq:HC}
S_\text{F}(T,\phi)=-(4/L)\int^{L/2}_{-L/2}dx\int^{\infty}_{0}dEN(x,E,T,\phi)
\times\\\left\{f(E)\ln[f(E)]+[1-f(E)]\ln[1-f(E)]\right\},\nonumber
\end{align}
 and
$f(E)=\left\{1+\exp[E/T]\right\}^{-1}$ is the
Fermi-Dirac quasiparticle distribution function at temperature
$T$.
\par
Since we employ a numerical solution, we have access to study the
full proximity effect regime and also, in principle, an arbitrary
spatial modulation $h=h(x)$ of the exchange field. This is
desirable in order to clarify effects associated with non-uniform
ferromagnets, such as the presence of Bloch domain walls. \par In
this paper, we will consider two different types of magnetization textures:
homogeneous magnetization and a Bloch domain wall structure as shown
in the Fig. \ref{fig:model}\textit{i)} and \textit{ii)}, respectively. It
is seen from Fig. \ref{fig:model} part \textit{ii)} that for a domain
wall structure, the magnetic moment has two components unlike the
homogeneous type. The Bloch model is given by
$\mathbf{h}=h(\cos\theta \hat{y}+\sin\theta \hat{z})$ and its
structure is shown in Fig. 1 part \textit{ii)}, where we
defined $\theta=-2\arctan(x/d_{W})$\cite{Konstandin},
with $d_{W}$ as the width of domain wall. Moreover, the center of the F layer is located at the origin $x=0$. Below, we shall consider a domain wall of width $d_W/d_F=0.5$, thus ensuring that the magnetization is fully aligned with the $z$-axis at the interfaces.

\begin{figure*}
\includegraphics[width=18cm]{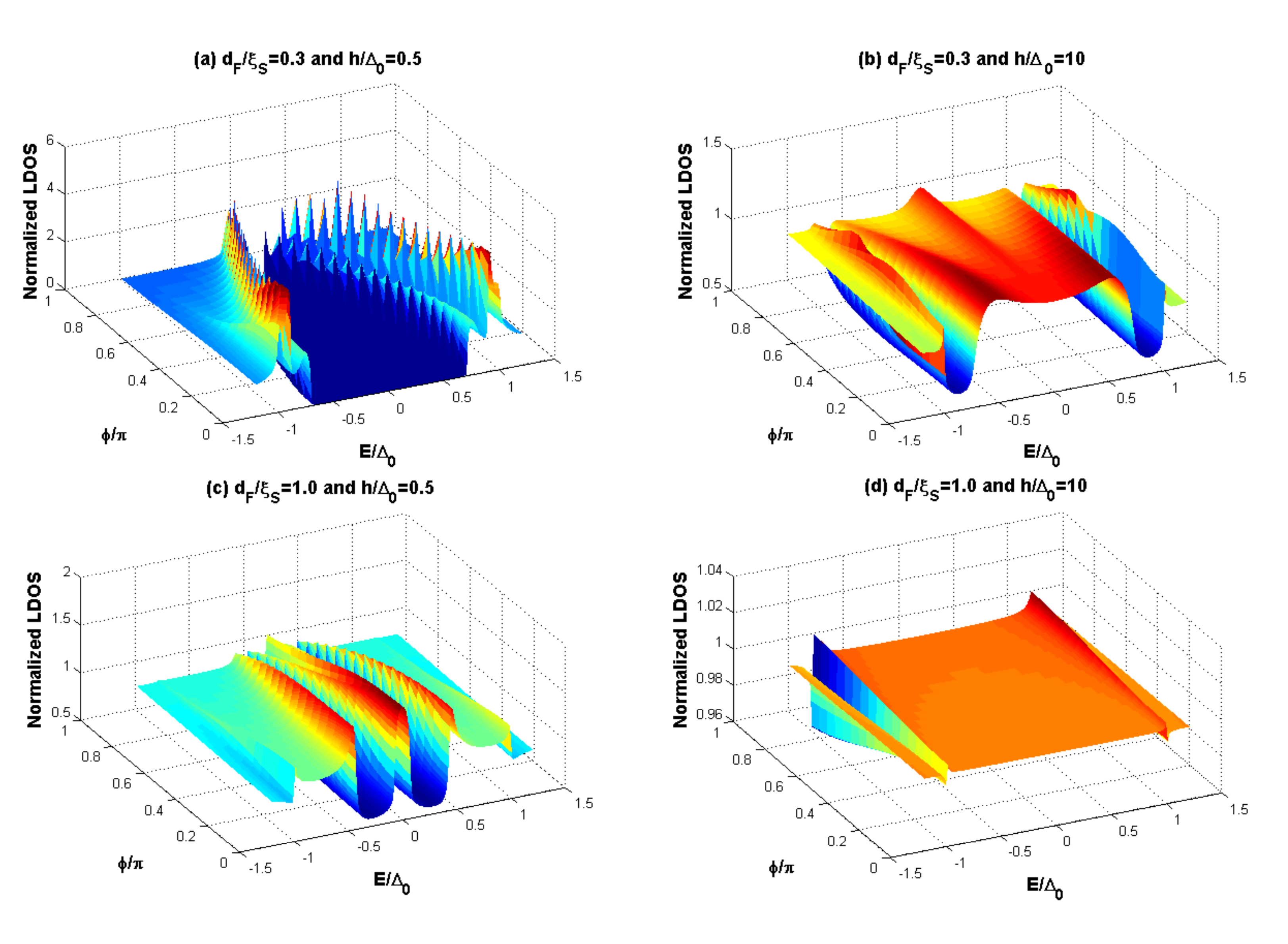}
\caption{\label{fig:DOS2} (Color online)The
normalized local density of states vs. energy and phase difference
between two $s$-wave superconducting leads and inhomogeneous
magnetization texture for several values of thickness of
ferromagnetic layer $d_{F}/\xi_{S}$ and exchange field
$h/\Delta_{0}$. The DOS is evaluated in the middle of the junction, \textit{i.e.} $x=0$.}
\end{figure*}

\begin{figure*}
\includegraphics[width=18cm]{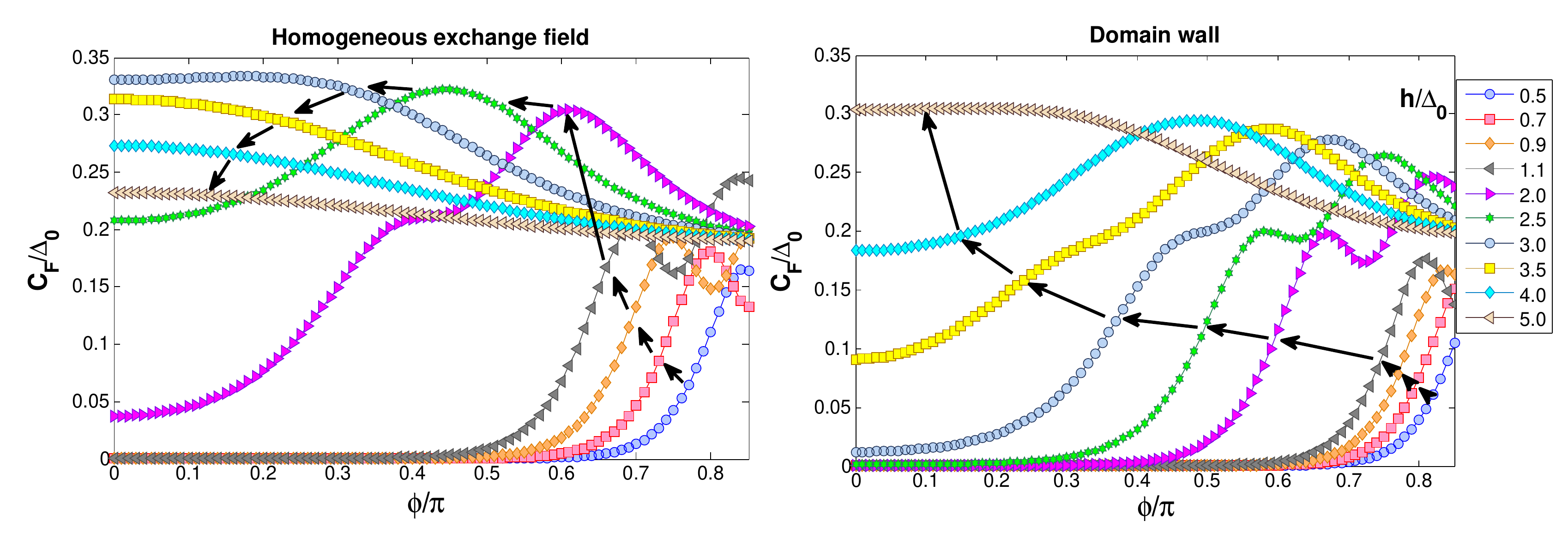}

\caption{\label{fig:SH} (Color online) The normalized electronic
specific heat of the diffusive S$\mid$F$\mid$S josephson junction
vs. phase difference between two $s$-wave supperconducting leads
with a ferromagnetic layer featuring a homogeneous (left panel)
and inhomogeneous Bloch domain wall magnetization texture (right
panel). The arrows indicate an increasing exchange field $h$.}
\end{figure*}

\section{Results and Discussion}
In this section, we present our main results of the paper, namely
the manner in which the DOS of a S$\mid$F$\mid$S diffusive Josephson
junction is altered in the presence of phase difference between two $s$-wave
superconducting leads and consequently the behavior of the electronic specific
heat of such junctions. We consider a
ferromagnet with two types of magnetization texture: homogenous and inhomogeneous (Bloch domain walls), including also the role of spin-active interfaces at the two
boundaries. In the quasiclassical framework employed here, we have to consider an exchange field much weaker than the Fermi energy in order to remain within the regime of validity. For a weak, diffusive
ferromagnetic alloy such as PdNi, the exchange field
$h$/$\Delta_{0}$ can be varied from a few meV to tens of meV by changing the relative content of Pd and Ni. Even weaker exchange fields
$h$ of order meV are found in for instance Y$_4$Co$_3$,
Y$_9$Co$_7$, and TiBe$_{1.8}$Cu$_{0.2}$ \cite{kolo_jpf_84}. Therefore,
we shall here consider exchange fields ranging from $0.5$ meV up to
$5$ meV. The scenario of a thin junction $d_{F}/\xi_{S}$=$0.3$ will be contrasted with that of a thick junction $d_F/\xi_S=1.0$.
For a superconducting lead like Nb with a coherence length of
$\xi_{S}\simeq18$ nm, the ratio of $d_{F}/\xi_{S}$=$0.3$ provides a ferromagnetic layer
thickness equal to $6$ nm which is experimentally accessible \cite{oboznov_prl_06}. The temperature will be
fixed at $T$=$0.05T_{c}$ and consequently our results are valid
for low temperature regime. The spin-dependent interfacial phase-shift (spin-DIPS)
term $G_{\phi}=G_{S}/G_{T}$ is obtained via the microscopic theory introduced
in the previous section, and depends \eg on the magnitude of the exchange
field and the interface transparency. Since we calculate $G_S$ microscopically for a simplified model with a Dirac tunneling barrier, $G_S$ is not treated as a phenomenological parameter here. We choose $\mu_F = 1$ eV and $\mu_S = 10$ eV for the
Fermi level in the ferromagnet and superconductor, respectively,
and consider a relatively low transparency barrier of $Z_0=3$. The
electron mass $m_F$ and $m_S$ in both of the F and S regions is taken
to be the bare one ($\simeq 0.5$ MeV). The ratio of the electronic
resistances of the barrier region and the ferromagnet layer is assumed to be
$\zeta$=$R_{B}/R_{F}$=$4$ throughout our computations. We also insert a small imaginary part
$\delta$=$10^{-3}$$\times$$\Delta_{0}$ into quasiparticle
energies \textit{i.e} $E \to E + \i\delta$ for
access to more stability in our computations. The small imaginary
part can be interpreted as accounting for inelastic scattering. As we discuss below,
we find that the specific heat of the S$\mid$F$\mid$S diffusive
junction can be strongly enhanced by changing the phase difference between
the two singlet superconducting leads from $0$ up to values near
$\pi$ for both a homogeneous exchange field and in the domain wall case. Due to limitations of our numerical code, we were not able to investigate phase differences $\phi$ very close to $\pi$. The huge enhancement of the
specific heat can be seen even for values of the exchange field several times the superconducting gap in the domain wall case.
Upon increasing the magnitude of the exchange
field further to values $h\gg\Delta_0$, this effect vanishes. The enhancement is most resilient towards an increase in $h$ in the case where a domain wall is present. Both the enhancement of the specific heat and its persistence in the domain wall case can be
understood by investigating the DOS in the ferromagnetic region. We now proceed to a presentation of our main results.

\subsection{Density of states (DOS) of a S$\mid$F$\mid$S junction at
low temperatures}

In this section, we discuss the behavior of
the DOS in a ferromagnetic region by changing the phase difference between two
superconducting leads connected to it. We fix the temperature at $T$=$0.05T_{c}$
and also use from microscopically values for spin-DIPS term in the
two boundaries. The results are shown in Fig. \ref{fig:DOS1} for the homogeneous exchange field case, while the domain wall scenario is demonstrated in Fig. \ref{fig:DOS2}. In both figures, we provide a contour-plot of the DOS in the middle of the F layer as a function of quasiparticle energy $E$ measured from Fermi level and the superconducting phase difference $\phi$.

Let us first consider the homogeneous case shown in Fig. \ref{fig:DOS1}(a)
for the case $d_{F}/\xi_{S}$=$0.3$ and $h/\Delta$=$0.5$. The most obvious feature is that a minigap-structure is induced in the low-energy regime close to the Fermi level, flanked by a peak structure below the gap and at the gap. The minigap is shown to close as the phase difference moves towards $\phi=\pi$, as is also the case for S$\mid$N$\mid$S junctions \cite{hammer_prb_07}. In Fig. \ref{fig:DOS1}(c), the junction thickness is increased to $d_F/\xi_S=1.0$, and it is seen that the peak structures remain. The main difference from (a) is that the low-energy DOS is enhanced, indicating the odd-frequency correlations are present and comparable in magnitude to the even-frequency correlations. The minigap is split into two and is seen to shift away from zero energy. The appearance of the multiple peak structures as a function of energy $E$ originates from an effective superconducting gap felt by each spin species which is different in magnitude for spin-$\uparrow$ and spin-$\downarrow$ quasiparticles. This is similar to the scenario of thin-film superconductors subjected to an in-plane external magnetic field \cite{meservey}. When the exchange field is increased to $h/\Delta_0=10$ as shown in (b) and (d), only the standard BCS-coherence peaks remain at the gap, although much weaker in magnitude than in the bulk case.

We now turn to the domain wall case, shown in Fig. \ref{fig:DOS2}. The most noteworthy change from Fig. \ref{fig:DOS1} is that the zero-energy DOS is enhanced in (b) and (c). This observation signals that odd-frequency correlations are stronger in the domain-wall case, a finding which agrees with the results in Ref. \cite{linder_prb_09}. The physical reason for this is that the inhomogeneous magnetization texture generates not only the $S_z=0$ triplet component, but also the long-ranged $S_z=\pm1$ triplet components, which also are odd in frequency due to the isotropization caused by the impurity scattering.

The presentation of the DOS and its dependence on the energy $E$ and superconducting phase difference $\phi$ presented in this section is a useful preliminary which, as we shall see, explains the origin behind our main result of a strongly enhanced specific heat, which we shall now move on to.

\subsection{Electronic specific heat of S$\mid$F$\mid$S junction at low temperatures}

Let us consider the electronic specific heat of the S$\mid$F$\mid$S diffusive
Josephson junction vs. phase difference of the two superconducting
leads. As mentioned in the Introduction, the phase difference is an experimentally tunable quantity by means of \eg current-biasing the junction or applying an external magnetic field in a SQUID-like geometry. Since the proximity effect is in general much weaker in the low-energy regime for $d_F/\xi_S=1.0$, we focus here on the more interesting case $d_{F}/\xi_{S}$=$0.3$.

\begin{figure}[b!]
\includegraphics[width=8cm]{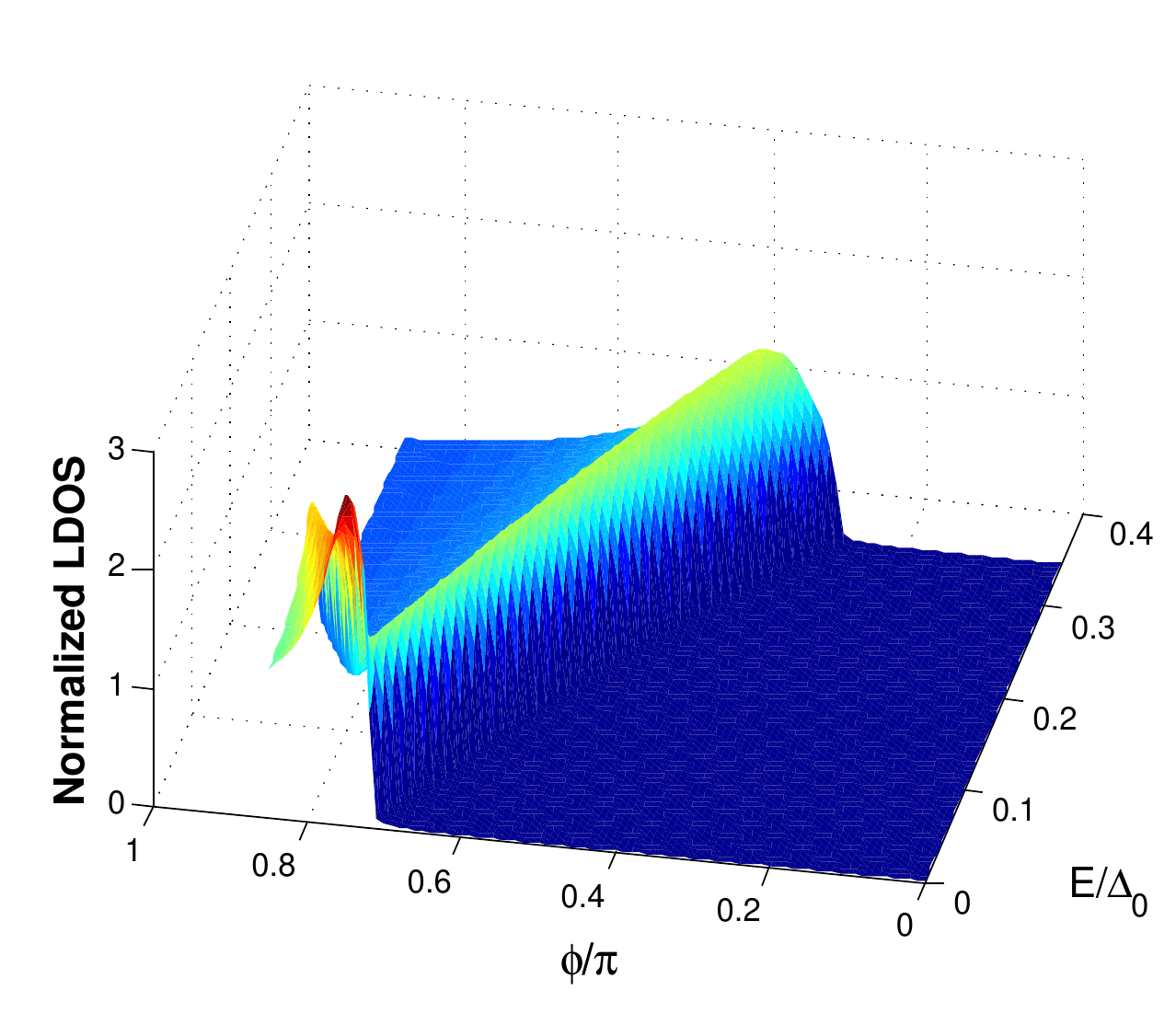}
\caption{\label{fig:DOS3} (Color online) The
normalized local density of states of the diffusive
S$\mid$F$\mid$S junction vs. phase difference and energy a for homogeneous
structure of ferromagnetic layer with exchange field
$h/\Delta_{0}=1.1$ and thickness $d_F/\xi_S=0.3$.}
\end{figure}

The results are shown in Fig. \ref{fig:SH}. As seen, the left
panel is related to the homogeneous exchange field scenario while the right panel is related to
the inhomogeneous magnetization in the form of a Bloch domain wall. In both cases, the curves show a
giant enhancement of the normalized specific heat when the exchange field is comparable in magnitude to the superconducting gap. For larger exchange fields, the specific heat becomes a monotonic, nearly constant function of the phase difference $\phi$. We note that the enhancement persists for larger values of $h$ in the domain wall case (up to $h/\Delta_0\simeq3.0$) compared to the homogeneous case. The physical reason behind the enhancement of the specific heat stems from the dependence of the DOS on $\phi$, as shown in Figs. \ref{fig:DOS1} and \ref{fig:DOS2}. For instance, for a very weak exchange field $h/\Delta_0=0.5$, the DOS-plots presented in the previous section showed how the minigap closed with increasing $\phi$. Since it is the low-energy DOS that mainly contributes to the electronic specific heat, increasing the phase difference $\phi$ will naturally lead to an increase in $C_F$. More specifically, we have verified numerically that at $T/T_c$=$0.05$, only energies up to $E/\Delta_0\simeq0.35$ contribute to the specific heat integral in Eq. (\ref{eq:HC}).

By increasing the magnitude of the exchange field in the
ferromagnetic layer, a kink appears in the specific heat. To identify the cause of the
appearance of the kinks, one should investigate the related DOS of
the system. Consider now for concreteness the DOS in the homogeneous exchange field case with
 $h/\Delta_{0}$=$1.1$, which is seen to display a kink in the specific heat in the left panel of Fig. \ref{fig:SH}. The kink of this curve
appears near $\phi/\pi\simeq0.7$, consequently leading us to plot
the DOS of the system near this value vs. $E/\Delta_{0}$ and
$\phi/\pi$. The resulting DOS is shown in Fig. \ref{fig:DOS3}. As
seen, the cause of appearance of a kink near $\phi/\pi\simeq0.7$
is the zero-energy peak that occurs in this region of the phase
difference. Such a zero-energy peak should be a direct result of
the manifestation of odd-frequency correlations in the system
\cite{oddfreq}. In Fig. \ref{fig:DOS3}, it is seen that an abrupt
conversion takes place at $\phi/\pi\simeq0.7$ along the $E=0$ line
from a fully suppressed DOS to an enhanced value compared to the
normal-state. Such an abrupt conversion was also very recently
studied in Ref. \cite{linder_prl_09}, where it was demonstrated
that the conversion was associated with a transition from pure
even-frequency to pure odd-frequency correlations. The
simultaneous decrease of the DOS when moving away from the Fermi
level results in a rapid decrease of the specific heat, thus
leading to the non-monotonic behavior shown in Fig. \ref{fig:SH}.

\section{Summary}

In summary, we have considered the density of states and
electronic specific heat of the diffusive S$\mid$F$\mid$S
Josephson junction both for a homogeneous
and inhomogeneous magnetization texture, including the role of spin-active interfaces.
We find that the electronic specific heat of the
S$\mid$F$\mid$S junction can be tuned to undergo a strong enhancement by increasing the
phase difference between two superconducting leads. The experimental requirement for observation of this effect is that the width $d_F$ of the ferromagnetic interlayer is considerably smaller than the superconducting coherence length (typically $d_F$ in the range 5-10 nm), and that the exchange field is comparable in magnitude to the gap. The effect persists in the domain wall case up to exchange fields $h/\Delta_0\simeq3$, yielding $h$ in the range 4-7 meV for a weak ferromagnetic alloy. Our prediction may be tested by \eg calorimetry measurements
of the Josephson junction, and the
results reported here could have interesting consequences for nanoscale devices relying
on an active tuning of their thermodynamic properties.

\acknowledgments I. B. Sperstad and T. Yokoyama are thanked for helpful
discussions. J.L. and A.S. were supported by the Research Council
of Norway, Grants No. 158518/432 and No. 158547/431 (NANOMAT), and
Grant No. 167498/V30 (STORFORSK).

\end{document}